\documentclass[10pt,conference]{IEEEtran}

\usepackage{amsmath,amssymb,url,tabularx}
\usepackage{graphicx,cite}
\usepackage{epstopdf}
\usepackage{color}

\newcommand{\D}{\displaystyle}
\newcommand{\R}{{\mathbb R}}

\begin{document}

\title{Anticipatory Radio Resource Management for Mobile Video Streaming with Linear Programming}

\author{\IEEEauthorblockN{Dimitrios Tsilimantos, Amaya Nogales-G\'omez, and Stefan Valentin}
	\IEEEauthorblockA{
		 Mathematical and Algorithmic Sciences Lab,\\  
		 FRC, Huawei Technologies\\ 
		 \{dimitrios.tsilimantos, amaya.nogales.gomez, stefan.valentin\}@huawei.com}}

\maketitle

\begin{abstract}
In anticipatory networking, channel prediction is used to improve communication performance. This paper describes a new approach for allocating resources to video streaming traffic while accounting for quality of service. The proposed method is based on integrating a model of the user's local play-out buffer into the radio access network. The linearity of this model allows to formulate a Linear Programming problem that optimizes the trade-off between the allocated resources and the stalling time of the media stream. Our simulation results demonstrate the full power of anticipatory optimization in a simple, yet representative, scenario. Compared to instantaneous adaptation, our anticipatory solution shows impressive gains in spectral efficiency and stalling duration at feasible computation time while being robust against prediction errors.
\end{abstract}

%%%%%%%%%%%%%%%%%%%
\section{Introduction}\label{sec:intro}
\bstctlcite{IEEEbibChanges:BSTcontrol}
%%%%%%%%%%%%%%%%%%%
Video streaming generated 45\% of all mobile data traffic in 2014 and is predicted to increase to 62\% by 2019 \cite{cisco2014global}. 
Although much effort has been spent to increase the capacity of mobile networks, it is still a major challenge for operators to assure sufficient streaming quality for mobile users. The technical difficulty here is to provide high resolution and fluency while reaching high Spectral Efficiency  (SE) with a time-variant wireless channel. Keeping the user's video play-out buffer filled at high rate, is often not possible or inefficient in difficult coverage situations, with interference, or at high speed. Once the play-out buffer runs empty, the video stalls, and the user's experience is heavily reduced. There is now a wide consensus that stalls are a major cause of dissatisfaction for the users of mobile streaming services \cite{HWwhitepaper,BufferBased,seufert15:HAS_QoE}.

\subsection{Idea and Contributions}
In this paper, we address this problem by Anticipatory Radio Resource Management (ARRM). Based on the knowledge of future channel states, the Base Station (BS) allocates wireless channel resources over upcoming time slots in order to fill the user's play-out buffer before a poorly covered area is reached. While moving through this area, the user perceives a fluent video stream from the preemptively filled buffer without using wireless channel resources. At the same time, these resources can be allocated to users with higher channel state. This increases overall spectral efficiency by reaching multi-user diversity gains, while satisfying the minimum bitrate requirement for streaming from the user's local memory but not from the channel.

Following this concept of ARRM, this paper contributes a model to predict the state of the user's play-out buffer at the BS. The linearity of this model allows to formulate two Linear Programming (LP) problems for ARRM. The first formulation maximizes spectral efficiency while avoiding stalls but becomes infeasible if the wireless channel state is too low to prevent a buffer under-run. Our second formulation avoids this problem by trading off stalling duration and SE. A detailed performance study shows that this formulation achieves outstanding SE gains at high QoS. These gains are reached at feasible computational time and decrease only slightly if channel prediction errors are taken into account.

The focus of this paper is entirely on the RRM for the final hop in mobile streaming. Aspects related to bottlenecks in the backbone and to the optimization of content storage (e.g., caching, CDN replication) are not considered in this work. Nonetheless, our model covers adaptive streaming, e.g., \cite{pantos11:hls_ietf,iso12:mpeg_dash}, by including a time-variant traffic rate in the optimization.

\subsection{Related Work}
The anticipatory, or proactive, allocation of wireless channel resources is typically used to compensate for delayed channel state information for a small number of upcoming transmission times \cite{ana05:chpre_schedulers,mehdi15:chpred_cqi_aging_lte}. Operating close to the coherence time, these schemes predict small-scale fading in the millisecond regime. Such short-term prediction is inapplicable for video streaming. This is a consequence of the relatively large segment size of common HTTP Adaptive Streaming Protocols, e.g., \cite{iso12:mpeg_dash}. To transfer a single segment, current cellular networks \cite{astely2009lte} may require hundreds of milliseconds transmission time. Thus, ARRM operates at a time scale where small-scale fading averages out and propagation loss dominates, which becomes time-variant with user mobility. As a result, channel prediction at this time scale is often based on combining the prediction of user trajectory with coverage data \cite{Yao2012,Riiser2012}. Based on analyzing coverage maps, a linear model for the prediction error was presented in \cite{SVLongTermPrediction}, which will be used in Section \ref{sec:results}.

Based on such long-term prediction of the wireless channel, only few authors have studied the anticipatory resource allocation for media streaming so far. This paper extends \cite{valentin13} by including initial buffer states, avoiding infeasible cases by trading off SE with QoS, and by studying robustness to prediction errors. In \cite{ValueOfKnowingTheFuture} an upper bound for ARRM was presented but only for the single-user case and without an operational formulation of the problem. Our paper goes beyond this work by presenting two tractable formulations for the multiple user case together with a concise study. Recently, a RRM framework was proposed in \cite{essaili15:RRM_HAS_LTE} which maximizes a non-linear Utility function by a greedy algorithm. Besides providing a linear formulation, our work exploits anticipation for the allocation of resources.

\subsection{Paper Structure}
The remainder of the paper is organized as follows. In Section \ref{sec:model}, we describe the resource allocation model for the multi-user case and discuss the proposed optimization problems. We present our simulation results in Section \ref{sec:results} and conclude with our remarks in Section \ref{sec:conclusion}.

\section{A Mathematical Model for Anticipatory RRM}
\label{sec:model}
In this paper we consider the downlink of a multi-user Orthogonal Frequency Division Multiple Access (OFDMA) system when the average channel gain is predicted for the video users over the next $T$ time slots, the prediction horizon. The OFDMA system is widely used in different standards such as in Long Term Evolution (LTE), \cite{astely2009lte}. In such a system, the bandwidth is divided into $N$ Physical Resource Blocks (PRBs), each one with a bandwidth $B$ which can be assigned to the different users. Let us define the set of users, time slots in the prediction horizon and BSs as $\mathcal{K}=\{1,\hdots,K\}$, $\mathcal{T}=\{1,\hdots,T\}$ and $\mathcal{M}=\{1,\hdots,M\}$ respectively. 

\subsection{Parameters and Variables}
We consider the following input parameters and variables:
\begin{itemize}
	\item Achievable data rate per PRB with $k\in\mathcal{K}, t\in\mathcal{T}$: 
	\begin{equation*}
		S_{k,t}=B\log_2\left ( 1+ \D\frac{P|\widehat{h}_{k,t}|^2}{\Gamma(\sigma ^2+I)} \right ),
	\end{equation*}
	where $P$ is the transmit power, $\sigma^2$ and $I$ is the noise and the interference respectively, $|\widehat{h}_{k,t}|^2$ is the predicted channel gain and $\Gamma$ is the Signal to Interference plus Noise Ratio (SINR) gap that accounts for the bit error rate in practical modulations. 
	\item Required play-out video bit rate $V_{k,t}$ with $k\in\mathcal{K}, t\in\mathcal{T}$. Note that the time index covers adaptive streaming, e.g., \cite{pantos11:hls_ietf,iso12:mpeg_dash}. We assume this value to be obtained by cross-layer signaling or traffic profiling.
	\item Base station assignment parameter with $k\in\mathcal{K}, t\in\mathcal{T}$:
	\begin{equation*}
		a^m_{k,t}=   \left\{\begin{array}{ll}
			1 & \mbox{if user } k \mbox{ is at time slot } t \mbox{ in BS } m  \\
			0 & \mbox{otherwise}
		\end{array} 
		\right. 
	\end{equation*}
\end{itemize}

In the following, we use the upper index $d$ for $V_{k,t}$ and $S_{k,t}$ when we want to refer to bits instead of rates, i.e. $V^d_{k,t}=V_{k,t}T_d$ and $S^d_{k,t}=S_{k,t}T_d$ respectively, where $T_d$ is the duration of one time slot. Moreover, we consider the following variables that characterize the problem:
\begin{itemize}
	\item Fraction of assigned resources $\omega_{k,t}\in\R^+$ with $k\in\mathcal{K}, t\in\mathcal{T}$ that represent the proportion of overall PRBs assigned to user $k$ at time slot $t$. 
	\item Buffer state with $k\in\mathcal{K}, t\in\{0\}\cup \mathcal{T}$:
	\begin{equation*}
		z_{k,0}=\zeta_k,\quad 0\leq z_{k,t}\leq Z_k,
	\end{equation*}
	where $z_{k,t}$ is the remaining data for user $k$ at the end of time slot $t$, $Z_k$ is the buffer size and $\zeta_k$ is the initial buffered data. 
	\item Stalling time $\ell_{k,t}$ with $k\in\mathcal{K},t\in\mathcal{T}$, which represents the fraction of time slot $t$ for which user $k$ did not receive the required play-out data. 
\end{itemize}

\subsection{LP Formulation Without Stalls}
\label{sec:ARRM-nostalls}
\begin{figure}[t]
	\centering
	\label{fig-downloaded}
	\includegraphics[width=0.325\linewidth]{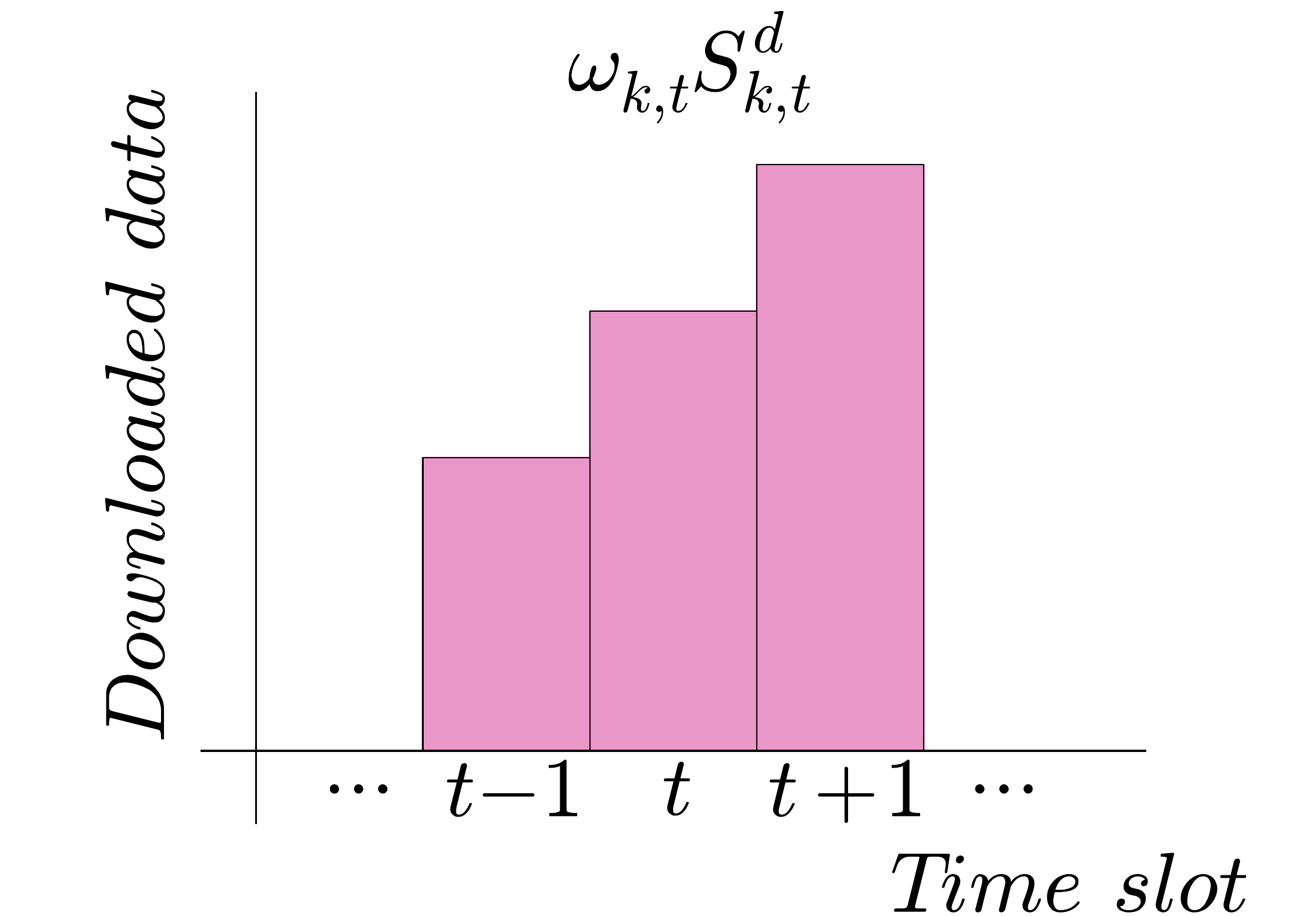}
	\centering
	\label{fig-played}
	\includegraphics[width=0.325\linewidth]{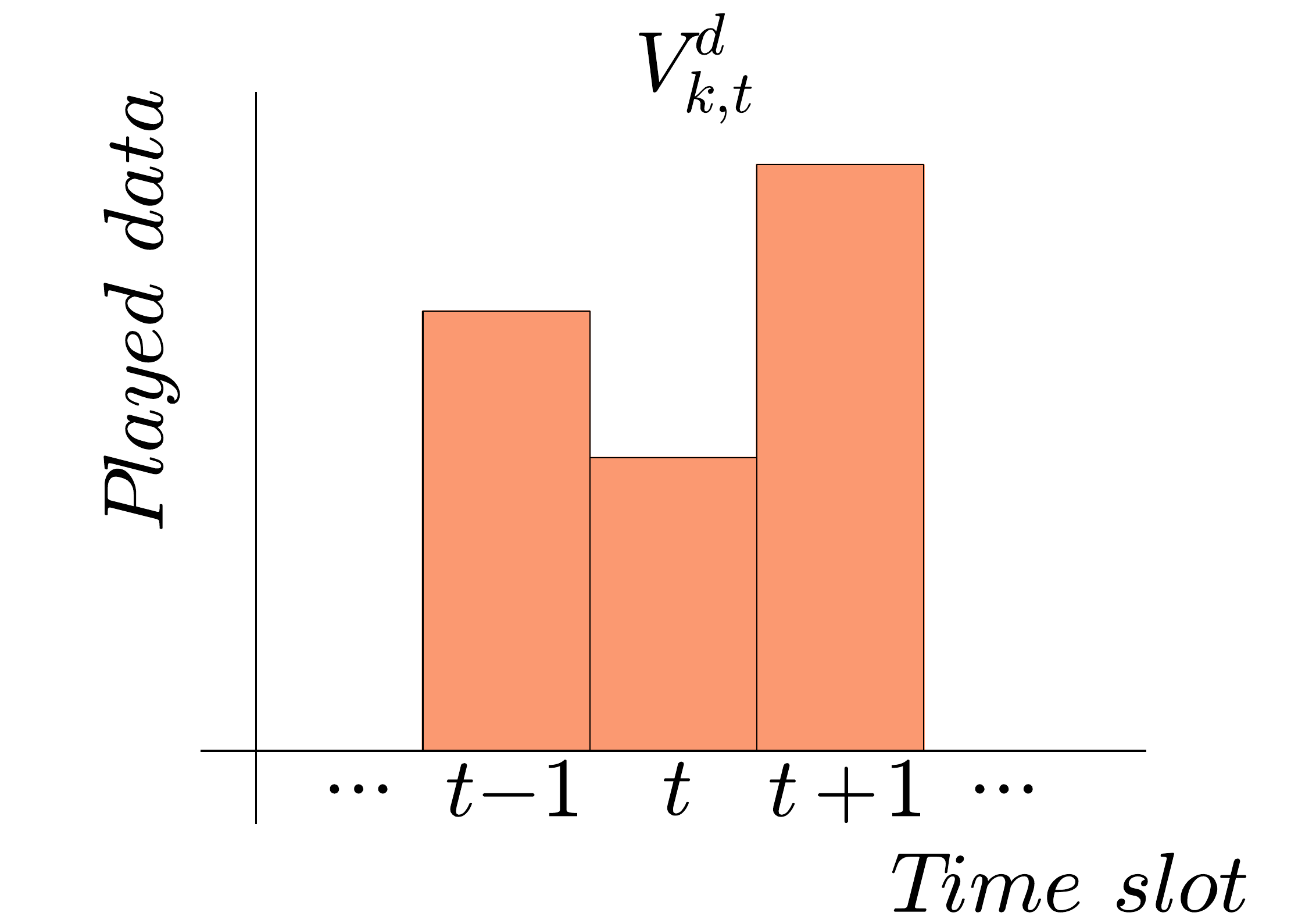}
	\label{fig-buffered}
	\includegraphics[width=0.325\linewidth]{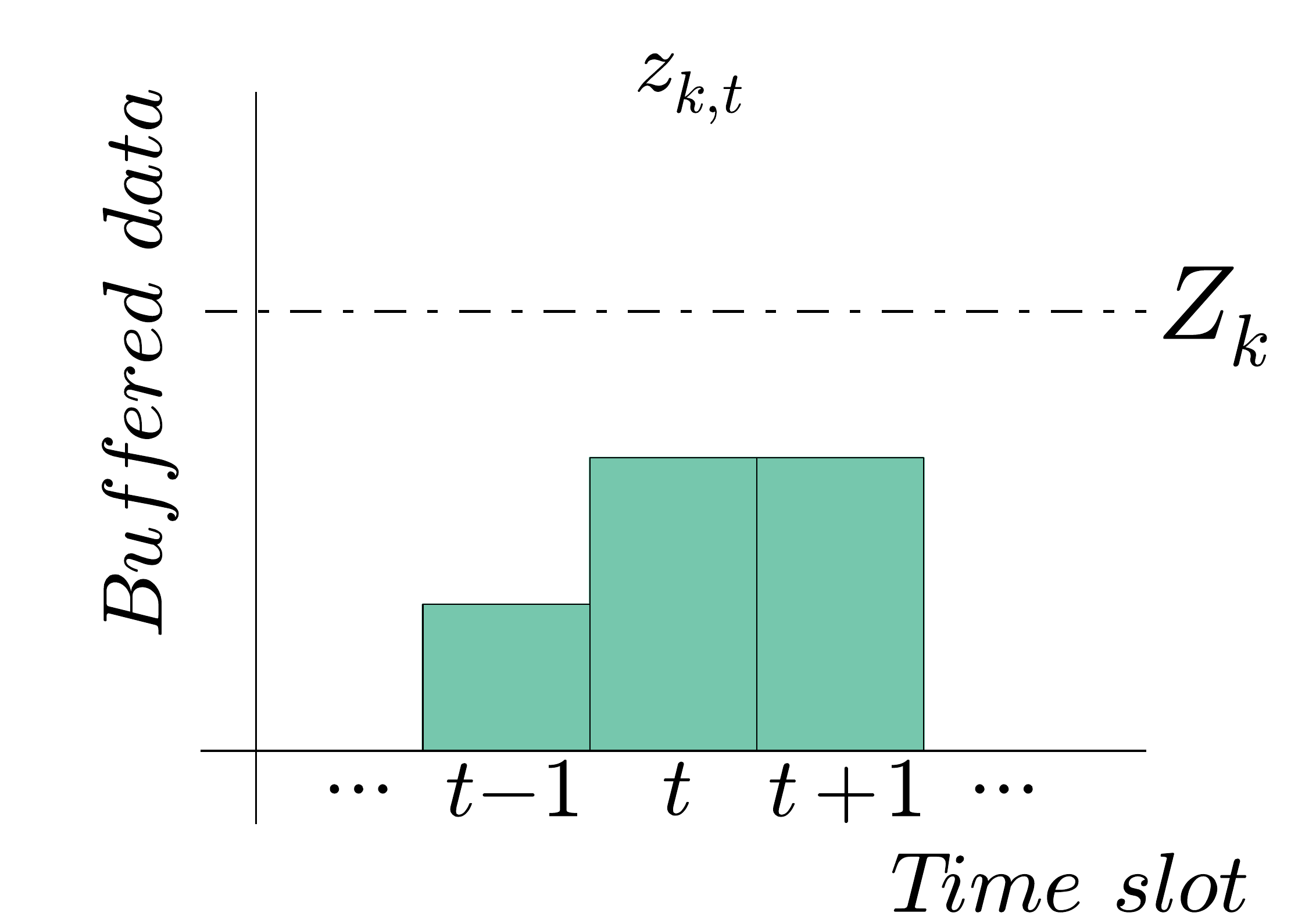}
	\caption{Model for the user's play-out buffer, illustrated for various time slots. The buffer state at the end of time slot $t$ is the result of the remaining buffered data from previous time slot (right) plus the downloaded data (left) minus the played data (center).}
	\label{fig:buffer}
\end{figure}

The proposed model is illustrated in Fig. \ref{fig:buffer}, which shows the evolution of the buffer state over the time slots. In general, each user $k$ requires $V^d_{k,t}$ bits per time slot $t$ in the buffer to play a video. At any time slot for a given user, there are $\omega_{k,t}S^d_{k,t}+z_{k,t-1}$ bits in the buffer, and a stall will take place if this amount is less than $V^d_{k,t}$. On the contrary, if $\omega_{k,t}S^d_{k,t}+z_{k,t-1}\geq V^d_{k,t}$, the unused bits are carried over to the next time slot.
To avoid stalls, we formulate the following constraints for the proposed buffer model:
\begin{IEEEeqnarray*}{rCl}
		z_{k,1}&=&\omega_{k,1}S^d_{k,1}+\zeta_k-V^d_{k,1}\vspace{-5pt}\\
		\vspace{-5pt}
		z_{k,2}&=&\omega_{k,2}S^d_{k,2}+z_{k,1}-V^d_{k,2}=\sum_{t=1}^2\left(\omega_{k,t}S^d_{k,t}-V^d_{k,t}\right)+\zeta_k \vspace{-10pt}\\
		\vspace{-10pt}
		&\vdots&  \\
		z_{k,T}&=&\sum_{t=1}^T\left(\omega_{k,t}S^d_{k,t}-V^d_{k,t}\right)+\zeta_k ,
\end{IEEEeqnarray*}
which can be reduced to the constraint $\forall k\in\mathcal{K}, \forall t\in\mathcal{T}$:
\begin{IEEEeqnarray}{rrl}
	\label{buffer-1}
	z_{k,t}=\D\sum_{i=1}^t\left(\omega_{k,i}S^d_{k,i}-V^d_{k,i}\right)+\zeta_k,&\;\;&z_{k,t},\omega_{k,t}\in\R^+.
\end{IEEEeqnarray}

Based on this linear buffer model, we can formulate the previously-stated resource allocation problem for multiple video users as
\begin{subequations}
	\begin{equation}
	\label{LP-nostalls-obj}
	\min_{\omega}\,\D\sum_{k\in\mathcal{K}}\D\sum_{t\in\mathcal{T}}\omega_{k,t}  \vspace{-5pt}
\end{equation}
\quad\quad\quad s.t. \hfill \vspace{-5pt}
\begin{IEEEeqnarray}{rCl}
	\label{LP-nostalls-1}
	0\leq \D\sum_{i=1}^t\left(\omega_{k,i}S^d_{k,i}-V^d_{k,i}\right)+\zeta_k \leq Z_k &\quad &\forall k\in\mathcal{K}, \forall t\in\!\mathcal{T},\\
	\label{LP-nostalls-2}
	\sum_{k\in\mathcal{K}}\omega_{k,t}\,a^m_{k,t}\leq N_m  &\quad &\forall t\in\mathcal{T},\forall m\!\in\!\mathcal{M}, \IEEEeqnarraynumspace\\
	\label{LP-nostalls-3}
	\omega_{k,t}\in\R^+ &\quad & \forall k\in\mathcal{K}, \forall t\in\!\mathcal{T},
\end{IEEEeqnarray}
\end{subequations}
where the objective function is chosen to minimize the total allocated resources, subject to the constraints for the buffer level and the available resources per BS. Since both the objective function and the constraints are linear, our formulation represents an LP, which can be solved with conventional optimization software \cite{CPLEX126} in polynomial time on the average.

\subsection{LP Formulation With Stalls}
\label{ARRM-stalls}
The previous optimization problem takes advantage of channel state prediction in order to maximize spectral efficiency while perfectly avoiding stalls. However, with consistently poor radio coverage, stalls will be unavoidable and problem \eqref{LP-nostalls-obj}-\eqref{LP-nostalls-3} will become infeasible. In this section, we propose a variant of the model described in Section \ref{sec:ARRM-nostalls} where stalls are allowed, but stalling time is minimized. First, we formulate the following constraints for the proposed buffer model:
\begin{subequations}
	\begin{IEEEeqnarray}{rCl}
		\label{model-stalls-1}
		z_{k,1}&=&V^d_{k,1}\ell_{k,1}+\omega_{k,1}S^d_{k,1}+\zeta_k-V^d_{k,1}\\
		\vspace{-3pt}
		z_{k,2}&=&V^d_{k,2}\ell_{k,2}+\omega_{k,2}S^d_{k,2}+z_{k,1}-V^d_{k,2} \vspace{-5pt}\\
		\vspace{-9pt}
		\nonumber 
		&\vdots&  \\
		\label{model-stalls-last}
		z_{k,T}&=&\sum_{t=1}^T\left(V^d_{k,t}\ell_{k,t}+\omega_{k,t}S^d_{k,t}-V^d_{k,t}\right)+\zeta_k ,
	\end{IEEEeqnarray}
\end{subequations}
which can be reduced to the constraint $\forall k\in\mathcal{K}, \forall t\in\mathcal{T}$:
\begin{IEEEeqnarray*}{rrl}
	\label{buffer-t}
	z_{k,t}=\D\sum_{i=1}^t\left(V^d_{k,i}\ell_{k,i}+\omega_{k,i}S^d_{k,i}-V^d_{k,i}\right)+\zeta_k,		\vspace{-9pt}\\
	\vspace{-3pt}
	z_{k,t},\omega_{k,t},\ell_{k,t}\in\R^+.
\end{IEEEeqnarray*}

In the previous model (\ref{LP-nostalls-obj})-(\ref{LP-nostalls-3}), $\omega_{k,t}S^d_{k,t}+z_{k,t-1}<V^d_{k,t}$ implies that the model is infeasible since it leads to $z_{k,t}<0$. Under the same circumstances, the new buffer model, however, gives $z_{k,t}=0$ and, thus, leads to a feasible solution. In addition, it yields a positive value for the stalling time $\ell_{k,t}$.
On the contrary, in the case that $\omega_{k,t}S^d_{k,t}+z_{k,t-1}>V^d_{k,t}$, the buffer model \eqref{model-stalls-1}-\eqref{model-stalls-last} could lead to unrealistic solutions where $z_{k,t}>0$ and $\ell_{k,t}>0$. We can avoid such cases by minimizing the stalling time in the objective function, which leads to the following solutions $\forall k\in\mathcal{K}, \forall t\in\mathcal{T}$:
\begin{IEEEeqnarray*}{rCl}
	\label{newbuffer-1}
	z_{k,t}&=&\max(\omega_{k,t}S^d_{k,t}+z_{k,t-1}-V^d_{k,t},0),\\
	\label{newbuffer-2}
	\ell_{k,t}&=&\D\frac{1}{V^d_{k,t}}\max(-\omega_{k,t}S^d_{k,t}-z_{k,t-1}+V^d_{k,t},0).
\end{IEEEeqnarray*}

Finally, we choose the objective function of this new model in order to minimize a trade-off between the total allocated resources and the stalling time. Controlling this trade-off requires to introduce the free parameter $\gamma \in\R^+$, where higher values of $\gamma$ prioritize stalling time minimization. Thus, we can formulate the previously stated resource allocation problem for multiple video users as the LP problem:

\begin{subequations}
	\begin{small}
		\begin{equation}
		\label{LP-stalls-final-obj}
		\min_{\omega,z,\ell}\, \D\sum_{k\in\mathcal{K}}\D\sum_{t\in\mathcal{T}} \left(\omega_{k,t} + \gamma \ell_{k,t}\right)
		\vspace{-5pt}
		\end{equation}
		\quad\quad\quad s.t. \hfill \vspace{-5pt}
		\begin{IEEEeqnarray}{rCL}
			\label{LP-stalls-final-2}
			\D\sum_{i=1}^t\left(V^d_{k,i}\ell_{k,i}+\omega_{k,i}S^d_{k,i}-V^d_{k,i}\right)+\zeta_k \geq0 &\enskip &\forall k\!\in\!\mathcal{K}, \forall t\!\in\!\mathcal{T},\IEEEeqnarraynumspace\\
			\D\sum_{i=1}^t\left(V^d_{k,i}\ell_{k,i}+\omega_{k,i}S^d_{k,i}-V^d_{k,i}\right)+\zeta_k \leq Z_k &\enskip &\forall k\!\in\!\mathcal{K}, \forall t\!\in\!\mathcal{T},\IEEEeqnarraynumspace\\
			\label{LP-stalls-final-3}
			\sum_{k\in\mathcal{K}}\omega_{k,t}\,a^m_{k,t}\leq N_m  &\enskip &\forall t\!\in\!\mathcal{T},\forall m\!\in\!\mathcal{M},\\[-5pt]
				\label{LP-stalls-final-5}
			\omega_{k,t},\ell_{k,t}\in\R^+  & \enskip& \forall k\!\in\!\mathcal{K}, \forall t\!\in\!\mathcal{T}. 
		\end{IEEEeqnarray}  
	\end{small}  
\end{subequations} 
\vspace{-15pt}
%%%%%%%%%%%%%%%%%%%
\section{Numerical Results}\label{sec:results}
%%%%%%%%%%%%%%%%%%%
We begin this section by discussing the performance metrics and simulation assumptions for the study of different resource allocation schemes. Then, we provide numerical results for the LP formulation \eqref{LP-stalls-final-obj}-\eqref{LP-stalls-final-5} for multiple users with video streaming traffic.

\subsection{Performance Metrics}    
We focus on the following three performance metrics:
\begin{enumerate}
	\item \textit{Cell spectral efficiency} (bits/s/Hz/cell): This critical measure for wireless networks is commonly defined as the data rate that the BS transmits over a given bandwidth, divided by the number of cells. According to our mathematical model, SE is given by
	\begin{equation}
		\label{eq: SE}
		SE = \frac{\sum_{k\in\mathcal{K}}\sum_{t\in\mathcal{T}_S} \omega_{k,t}S_{k,t}}{M B\sum_{k\in\mathcal{K}}\sum_{t\in\mathcal{T}_S} \omega_{k,t}},
	\end{equation}
	where $\mathcal{T}_S=\lbrace1,...,T_S\rbrace$ is the set of $T_S$ time slots until $K$ users are served.   
	    
	\item \textit{Stalling duration} (s): Since user mobility leads to time-variant channel gains, the allocated data rate may not always support the traffic rate of the video stream. If, in this case, the user's play-out buffer runs empty, the video stream stalls. Since stalling duration is recognized as a main factor for QoS \cite{HWwhitepaper,BufferBased,seufert15:HAS_QoE}, we assign higher priority to the stalling duration than to the allocated resources. In mathematical terms, the parameter $\gamma$ in \eqref{LP-stalls-final-obj} is chosen high enough to ensure that the solution of the optimization gives the minimum resources for the minimum feasible stalling time.      
	\item \textit{Computational time} (s): Since RRM has to adapt to the state of channel and traffic in real time, the optimization problem has to be solved sufficiently fast. However, it is worth mentioning that the optimization does not have to be performed for each Transmission Time Interval (TTI). This results from the fact that HTTP-based streaming protocols separates the video stream into segments, each containing several seconds of video time. As only complete segments can be played, which requires typically hundreds of milliseconds transmit time per segment, it is sufficient to choose a slot duration $T_d$ at this time scale.
\end{enumerate}

%%%%%%%%%%%%%%%%%%% 
\subsection{Simulation Scenario and Parameters}
%%%%%%%%%%%%%%%%%%%
A simple two-cell scenario is assumed throughout the remainder of the paper. Video users arrive at the system following a Poisson arrival process with rate $\lambda=\frac{K}{T_NT_d}$, where $T_N\leq T_S$ is the number of time slots that the user stays connected. Each user moves with constant speed in a straight line from the first BS to the second one and requests video streaming, approximating the situation for vehicular users in a highway. The chosen inter-site distance corresponds to an typical LTE deployment in urban areas. Note that this simple scenario already captures the main idea of anticipatory resource allocation, since the channel gain in the cell edge between the two BSs can likely lead to stalls if multiple users are sharing the BS resources at the same time. By predicting the channel gain, the user's play-out buffer can be filled in advance under better channel conditions and be consumed in the cell edge, where the channel gain is low.

To account for the channel, we adopt the 3GPP path-loss model $\text{PL}=128.1+37.6\log_{10}d+L_s$ \cite{3GPP}, where $d$ (km) is the distance between the user and the serving BS, which we consider to be the nearest one, and $L_s$ is the shadowing factor.  For the sake of simplicity, we do not explicitly calculate interference and only a margin that includes both intra- and inter-cell interference is introduced. The SINR gap between the achieved spectral efficiency and the Shannon channel capacity is modeled as a simple function of Bit Error Rate (BER), as in \cite{Goldsmith}. Fast fading is not taken into account, as we assume that it is averaged out over hundreds of milliseconds of a time slot. A new optimization is performed either when a new user arrives at the system or after $T_c$ time slots, where $T_c\leq T$ is a fixed optimization step. This way, we keep the most recent results $\omega_{k,t}$ for each time slot and user, an approach which proves to have an important impact on the results for a specific range of values of $T$. 

Monte-Carlo simulations are performed and the average values are evaluated over 1000 iterations. A summary of the main simulation parameters are presented in Table \ref{table:SimulationParameters}, where the slot duration is $T_d=167$ ms given the respective values of user speed, inter-site distance and number of time slots $T_N$. Finally, the video traffic rates $V=1.5$ Mbits/s and $V=6$ Mbits/s are chosen, which are average rates for 720p and 2k HD streaming. The rest of the values of $V$ are used as intermediate levels. These values may seem high but are realistic for LTE metro-cells with a small number of video users. Our selection was confirmed by field measurements in two larger European cities and comes at no loss in generality. For simplicity, a common value $V$ is used for all users and time slots, although our formulation can handle varying traffic bitrates as well.
\begin{table}[!t]
	\renewcommand{\arraystretch}{1.3}
	\caption{Main simulation parameters.}
	\vspace{3pt}
	\label{table:SimulationParameters}
	\centering
		\begin{tabular}{|c||c|}
			\hline
			\bfseries Parameter & \bfseries Value\\
			\hline\hline
			Total BS Tx power $P$ & 46 dBm\\
			BS antenna gain & 18 dBi\\
			Available PRBs in a BS & 50 \\
			PRB bandwidth $B$ & 180 kHz\\
			Noise spectral density & -174 dBm/Hz\\
			Receiver noise figure & 10 dB\\
			Interference margin & 6 dB\\
			Shadow fading margin & 10 dB\\
			SINR gap $\Gamma$ & $-\ln\left(5\text{BER}\right)/1.5$ \cite{Goldsmith} \\
			BS inter-site distance & 500 m\\
			Total number of time slots $T_N$ & 100\\
			User speed & 30 m/s \\
			BS antenna height & 35 m\\
			Path-loss model  & 3GPP empirical \cite{3GPP}\\
			\hline
			Number of users $K$ & $\left[1,30\right]$\\
			Prediction horizon $T$ & $\left[1,T_N\right]$ time slots\\
			Maximum buffer size $Z_k$ & $20$ Mbits\\
			Video play-out rate $V$& $\lbrace1.5,2.5,4,6\rbrace$ Mbits/s\\
			Optimization step $T_c$ & $\left[1,T\right]$ time slots\\
			Trade-off parameter $\gamma$ & $\left[1,10^4\right]$\\
			\hline
		\end{tabular}
\end{table}

%%%%%%%%%%%%%%%%%%%
\subsection{Model of Channel Prediction Errors}\label{sec:Robustness}
%%%%%%%%%%%%%%%%%%%
Perfect channel prediction was assumed to be available so far in order to formulate the optimization problem in the previous sections. However, this is not the case in practical systems even with the most advanced predictors. Here we introduce a simple model for the channel prediction error $\epsilon_{k,t}$ that allows us to examine the robustness of our proposed algorithm. The predicted channel gain is then given by $|\widehat{h}_{k,t}|^2 = |h_{k,t}|^2 + \epsilon_{k,t}$ in dB scale, where $\epsilon_{k,t}$ is assumed to follow for all users a Normal distribution with zero mean and standard deviation that is linear to the prediction horizon, i.e. $\epsilon_{k,t}\sim N\left(0,\sigma^2_t\right)$, where $\sigma_t=\frac{t}{T}\sigma$ and $\sigma^2$ is the variance of the error for a prediction horizon of $T$. Although simple, this model incorporates the fact that the error increases with the prediction horizon and approximates well the linear gradient found in \cite{SVLongTermPrediction}. We leave more sophisticated analysis of prediction errors for our future work.

%%%%%%%%%%%%%%%%%%%
\subsection{Performance Study}\label{sec:PerformanceAnalysis}
\begin{figure}[!t]
	\centering
	\includegraphics[width=1\linewidth,trim={0cm 0 5cm 16cm},clip]{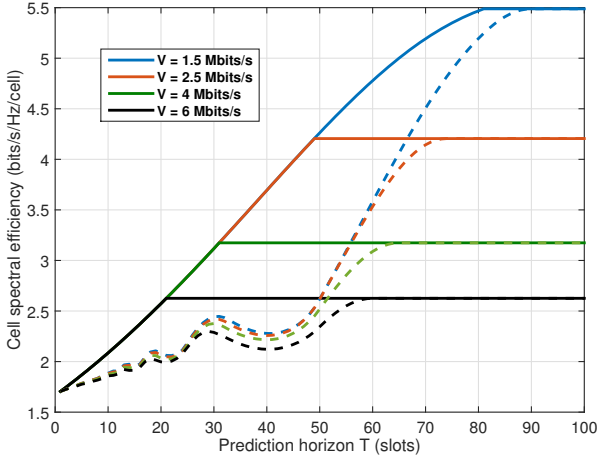}
	\caption{Single user SE as a function of $T$ and $V$; solid lines for  $T_c=1$ slot and dashed lines for  $T_c=T$.}
	\label{fig:OPovwr1PlusSeq_SE}
\end{figure}

%%%%%%%%%%%%%%%%%%%
We start by studying the effect of prediction horizon $T$ on the SE, under the assumption of perfect channel prediction. Fig. \ref{fig:OPovwr1PlusSeq_SE} shows the SE of a single user for a set of different video play-out rates $V$, as a function of the prediction horizon $T$. Two cases are examined, the first one when $T_c=T$, i.e. a new optimization is performed every $T$ slots, and the second one when $T_c=1$ slot.

We observe that as the encoding rate $V$ increases, lower SE is achieved above a certain value of $T$. This is due to the constraint of the maximum buffer size that allows for lower $V$ a longer video duration to be buffered close to the BS, where the channel conditions are better. Moreover, it is interesting to notice that as $T$ increases, the curves of the SE oscillate for $T_c=T$ until they reach a maximum constant value. Thus, in some cases, a higher prediction horizon leads to worse results in terms of SE. This counter-intuitive effect depends on the set of time slots for which the optimization is performed. For example, for $T=40$ slots, the second optimization is performed close to the cell edge where the resources are expensive in terms of bandwidth. On the other hand, for $T=30$ slots, the second optimization occurs early enough to anticipate for the cell edge and the buffer is then filled at a lower price. When a new optimization is realized every time slot for $T_c=1$ slot, then the curves are monotonic and the full potential of the prediction is exploited. This result shows that, even without prediction error, there is already a trade-off between computational effort and SE. We conclude that a small value for $T_c$ should be used in order to provide near to optimal results for all the different parameter values.

\begin{figure}[!t]
	\centering
	\includegraphics[width=1\linewidth,trim={0cm 0 1.5cm 13cm},clip]{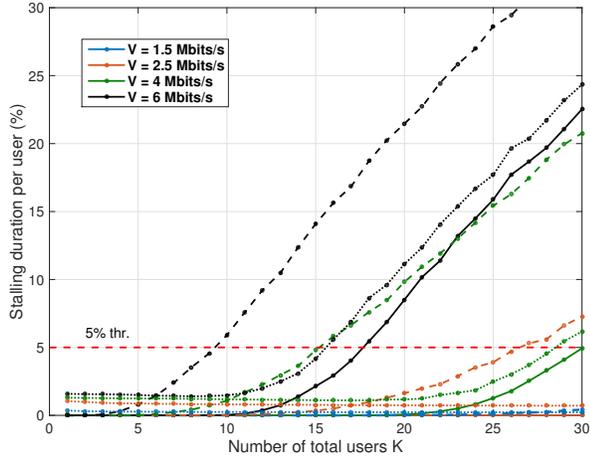}
	\caption{Stalling duration as a function of users $K$; solid lines for ARRM with perfect channel prediction, dotted lines for ARRM with $\sigma=10$ dB and dashed lines for baseline.}
	\label{fig:KV_Z20_StallsDuration}
\end{figure}

Now that we have illustrated the relation between basic parameters of our model, we proceed by studying the QoS in terms of stalling duration, with and without prediction errors. In the following figures, we denote by `ARRM' the case with $T=100$ and $T_c=20$ time slots respectively and by `baseline' the approach without prediction, where the BS instantaneously allocates the necessary resources to satisfy the given play-out data rate. Fig. \ref{fig:KV_Z20_StallsDuration} presents the stalling duration per user as a fraction of the total time the user spends in the system, for different values of $K, V$. As expected, ARRM (solid lines) provides a clear reduction of the stalling duration compared to the baseline (dashed lines). For example, by limiting the stalling duration to $5\%$, we can see that the maximum number of users, served under this QoS constraint, is almost doubled for $V=\lbrace4,6\rbrace$ Mbits/s. For lower $V$, the gains cannot be defined exactly in this example, since the QoS of ARRM is so high that more than the simulated number of $K=30$ users is supported. The effect of the channel prediction error is marginal as we can see for the case where ARRM with $\sigma=10$ dB (dotted lines). Here, the gains for $V=\lbrace4,6\rbrace$ Mbits/s are only slightly reduced compared to ARRM with perfect channel prediction.

\begin{figure}[!t]
	\centering
	\includegraphics[width=1\linewidth,trim={0cm 0 5cm 16cm},clip]{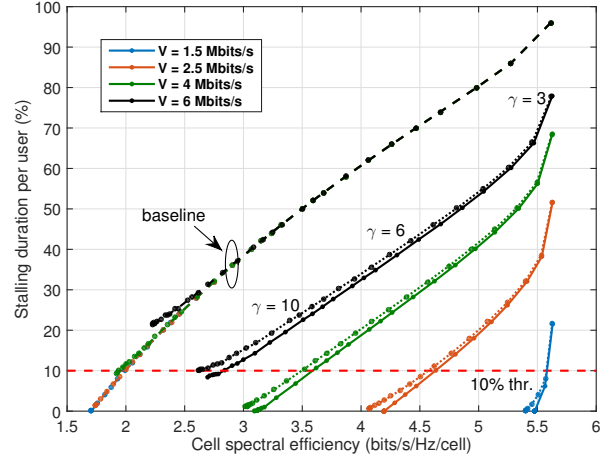}
	\caption{Cell SE vs. stalling duration for $K=20$;  solid lines for ARRM with perfect channel prediction, dotted lines for ARRM with $\sigma=10$ dB and dashed lines for baseline.}
	\label{fig:LP_tradeoff_Z20}
\end{figure}
\begin{figure}[!t]
	\centering
	\includegraphics[width=.8\linewidth]{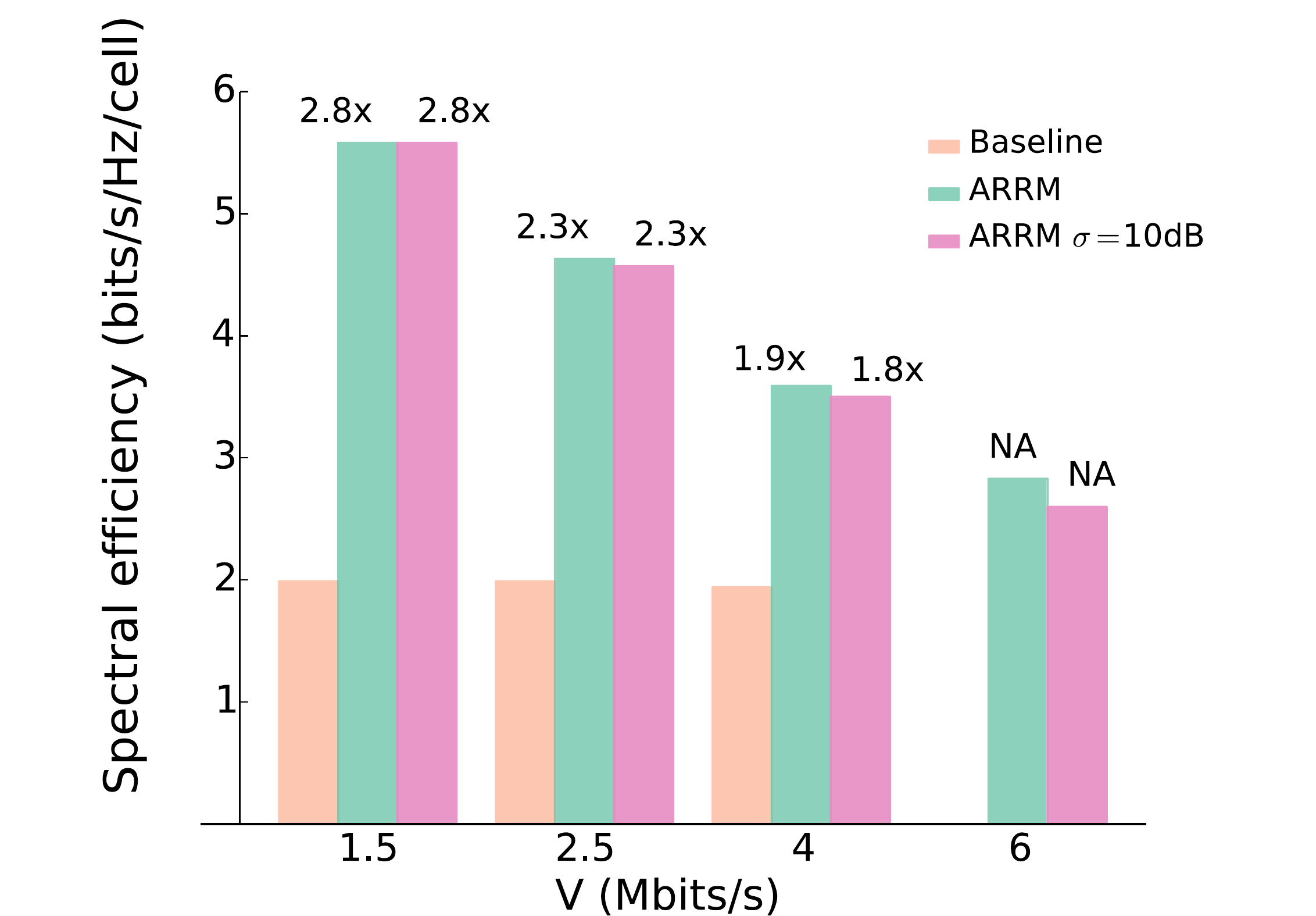}
	\caption{Cell SE for 10\% stalling duration and gains compared to baseline}
	\label{fig:bar-chart-stallDur-vs-SE}
\end{figure} 

Our ARRM formulation \eqref{model-stalls-1}-\eqref{model-stalls-last} trades off SE and stalling duration. In the previous results, a large value of $\gamma$ was assumed in order to prioritize the minimization of the stalling time. If $\gamma \leq\max\frac{V_{k,t}}{S_{k,t}}, \forall k,t$, then less resources can be allocated at the cost of higher stalling. The above expression of the $\gamma$ threshold is easily found by setting $\ell'_{k,t}=\ell_{k,t}+\delta$ and $\omega'_{k,t}=\omega_{k,t}-\frac{\delta V_{k,t}}{S_{k,t}}$, with $\delta>0$, that satisfy all the constraints of \eqref{LP-stalls-final-obj}-\eqref{LP-stalls-final-5} and then by verifying when this solution leads to a better objective value. Fig. \ref{fig:LP_tradeoff_Z20} illustrates the set of the optimal solutions for the complete range of $\gamma$ values from Table \ref{table:SimulationParameters} for $K=20$ users. We can see that ARRM achieves a better trade-off for all the curves, i.e. higher cell SE is obtained for a given value of stalling duration. For $V=6$ Mbits/s, we can also notice that the curves do not reach the x-axis, which means that although ARRM reduced stalling, stalls cannot be entirely avoided under such high load. The effect of the prediction error with $\sigma=10$ dB on the trade-off curves is higher as $\gamma$ increases, but remains marginal for all the studied cases. For a better illustration of the achieved SE gains, Fig. \ref{fig:bar-chart-stallDur-vs-SE} shows the SE reached at 10\% stalling duration. We can see in this figure that ARRM with $Z=20$ Mbits achieves an impressive increase of cell SE up to $2.8$ times, while satisfying the above QoS-constraint. For $V=6$ Mbits/s, we can also notice that the baseline does not satisfy the QoS-constraint. Note that these high gains are robust to channel prediction errors.

%%%%%%%%%%%%%%%%%%%
\subsection{Computational Time}\label{sec:CompTime}
%%%%%%%%%%%%%%%%%%%
We now study the computational time for the solution of the proposed LP formulation over different sets of parameters. Let us represent the number of simultaneously active users by the variable $K'$, with $K'\leq K$. We measure the computation time for one single optimization for different values of $K'$ and prediction horizon $T$ on a typical server processor (i.e., the Intel Xeon CPU running at $3.3$ GHz) using the optimization engine CPLEX v12.6, \cite{CPLEX126}.

The variables $K'$ and $T$ are the two key factors that define the size of the linear system in \eqref{LP-stalls-final-obj}-\eqref{LP-stalls-final-5} to be $2TK'$ variables and $T(2K'+M)$ constraints. Thus, $K'$ and $T$ have a large effect on the computational time. We have also verified that $V$ only insignificantly affects the runtime and assume, thus, a constant $V=1.5$ Mbits/s when studying this metric. Table \ref{table:MedianCompTime} provides the median computational time of one optimization. As we can see from this table, the computational time increases with the number of active users and with the prediction horizon. However, even for $T=100$ slots, the corresponding time strongly indicates that the proposed optimization problem can provide anticipatory resource allocation sufficiently fast for practical systems.
\begin{table}[!t]
	\renewcommand{\arraystretch}{1.5}
	\caption{Median computational time of one optimization.}
	\label{table:MedianCompTime}
	\centering
	\begin{footnotesize}
	\begin{tabular}{|c|c|c|c|c|}
		\hline
		$K'$&\multicolumn{4}{c|}{Time (ms)}\\
		\hline
		&Baseline& \multicolumn{3}{c|}{ARRM}\\
		\cline{3-5}
		& & $T=20$& $T=50$& $T=100$\\
		\hline
		\hline
		1& 0.06&0.23&0.44&0.93\\
		10&0.18&2.44&6.17&10.01\\
		20&0.24&6.17&15.67&23.26\\
		30&0.31& 11.48&28.61&42.74\\
		\hline
	\end{tabular}
\end{footnotesize}
\end{table}

%%%%%%%%%%%%%%%%%%%
\section{Conclusions}\label{sec:conclusion}
%%%%%%%%%%%%%%%%%%%
In this paper, we studied Anticipatory Radio Resource Management (ARRM) for mobile video streaming based on channel state prediction and knowledge of the traffic rate. An LP formulation was proposed that provides the optimal solution in a computationally efficient manner. 

Our numerical results for a representative scenario with multiple users and two base stations provide high insight. We verified that ARRM highly increases the QoS compared to the baseline scheme without anticipation. Further studying spectral efficiency and stalling time, displays the proper choice of the trade-off parameter and reveals an impressive spectral efficiency gain at high QoS. This gain is only slightly reduced if a model for channel prediction is introduced into the study. This shows how robust our ARRM formulation is against the side effects of practical channel prediction. Further practicality is demonstrated by a low computational time, which supports real-time solutions even for large instances of the problem.

In our future work, we will further study the robustness of ARRM under practical assumptions. This requires to study various error models for channel prediction and traffic rate estimation. We aim to include an error term into the ARRM formulation for further robustness. Finally, simulations of larger topologies, QoS, channel and traffic models are required, in order to verify the exceptional quality and efficiency in further scenarios.

\bibliography{IEEEabrv,ARRMbiblio}

% Generated by IEEEtran.bst, version: 1.14 (2015/08/26)
\begin{thebibliography}{10}
\providecommand{\url}[1]{#1}
\csname url@samestyle\endcsname
\providecommand{\newblock}{\relax}
\providecommand{\bibinfo}[2]{#2}
\providecommand{\BIBentrySTDinterwordspacing}{\spaceskip=0pt\relax}
\providecommand{\BIBentryALTinterwordstretchfactor}{4}
\providecommand{\BIBentryALTinterwordspacing}{\spaceskip=\fontdimen2\font plus
\BIBentryALTinterwordstretchfactor\fontdimen3\font minus
  \fontdimen4\font\relax}
\providecommand{\BIBforeignlanguage}[2]{{%
\expandafter\ifx\csname l@#1\endcsname\relax
\typeout{** WARNING: IEEEtran.bst: No hyphenation pattern has been}%
\typeout{** loaded for the language `#1'. Using the pattern for}%
\typeout{** the default language instead.}%
\else
\language=\csname l@#1\endcsname
\fi
#2}}
\providecommand{\BIBdecl}{\relax}
\BIBdecl

\bibitem{cisco2014global}
{Cisco Visual Networking Index}, ``Global mobile data traffic forecast update,
  2014--2019,'' \emph{white paper}, Feb. 2015.

\bibitem{HWwhitepaper}
Huawei, ``Mobile video service performance study,'' \emph{white paper}, Jun.
  2015.

\bibitem{BufferBased}
T.-Y. Huang \emph{et~al.}, ``{A Buffer-based Approach to Rate Adaptation:
  Evidence from a Large Video Streaming Service},'' in \emph{{SIGCOMM}}, Aug.
  2014, pp. 187--198.

\bibitem{seufert15:HAS_QoE}
M.~Seufert \emph{et~al.}, ``A survey on quality of experience of {HTTP}
  adaptive streaming,'' \emph{{IEEE} Commun. Surveys Tuts.}, vol.~17, no.~1,
  pp. 469--492, 2015.

\bibitem{pantos11:hls_ietf}
R.~Pantos and W.~May, ``{HTTP} live streaming,'' {IETF}, Informational
  Internet-Draft 2582, Sep. 2011.

\bibitem{iso12:mpeg_dash}
{ISO/IEC}, ``Dynamic adaptive streaming over {HTTP (DASH)},'' {ISO/IEC},
  International Standard DIS 23009-1.2, 2012.

\bibitem{ana05:chpre_schedulers}
A.~Aguiar, A.~Wolisz, H.~Lederer, and H.~Karl, ``Channel-adaptive schedulers
  with state-of-the-art channel predictors,'' in \emph{European Wireless}, Apr.
  2005, pp. 1--6.

\bibitem{mehdi15:chpred_cqi_aging_lte}
A.~Chiumento \emph{et~al.}, ``Adaptive {CSI} and feedback estimation in {LTE}
  and beyond: a {Gaussian} process regression approach,'' \emph{EURASIP Journal
  on Wireless Communications and Networking}, vol. 2015, no.~1, 2015.

\bibitem{astely2009lte}
D.~Ast{\'e}ly \emph{et~al.}, ``{LTE}: the evolution of mobile broadband,''
  \emph{IEEE Communications Magazine}, vol.~47, no.~4, pp. 44--51, 2009.

\bibitem{Yao2012}
J.~Yao, S.~S. Kanhere, and M.~Hassan, ``{Improving QoS in High-Speed Mobility
  Using Bandwidth Maps.}'' \emph{IEEE Trans.\ Mob.\ Comput.}, vol.~11, no.~4,
  pp. 603--617, 2012.

\bibitem{Riiser2012}
H.~Riiser \emph{et~al.}, ``{Video streaming using a location-based
  bandwidth-lookup service for bitrate planning.}'' \emph{{ACM Transactions on
  Multimedia Computing, Communications and Applications (ACM TOMCCAP)}},
  vol.~8, no.~3, p.~24, 2012.

\bibitem{SVLongTermPrediction}
Q.~Liao, S.~Valentin, and S.~Stanczak, ``Channel gain prediction in wireless
  networks based on spatial-temporal correlation,'' in \emph{SPAWC}, Stockholm,
  Sweden, June 2015, pp. 400--404.

\bibitem{valentin13}
\BIBentryALTinterwordspacing
S.~Sadr and S.~Valentin, ``Anticipatory buffer control and resource allocation
  for wireless video streaming,'' \emph{CoRR}, vol. abs/1304.3056, 2013.
  [Online]. Available: \url{http://arxiv.org/abs/1304.3056}
\BIBentrySTDinterwordspacing

\bibitem{ValueOfKnowingTheFuture}
Z.~Lu and G.~de~Veciana, ``{Optimizing stored video delivery for mobile
  networks: The value of knowing the future},'' in \emph{INFOCOM, 2013
  Proceedings IEEE}, Turin, Italy, April 2013, pp. 2706--2714.

\bibitem{essaili15:RRM_HAS_LTE}
A.~El~Essaili \emph{et~al.}, ``{QoE}-based traffic and resource management for
  adaptive {HTTP} video delivery in {LTE},'' \emph{{IEEE} Trans. Circuits Syst.
  Video Technol.}, vol.~25, no.~6, pp. 988--1001, Jun. 2015.

\bibitem{CPLEX126}
{IBM}-Cplex, {v. 12.6},
  \url{http://www-01.ibm.com/software/integration/optimization/cplex/}.

\bibitem{3GPP}
3GPP, ``{Further advancements for E-UTRA physical layer aspects},'' \emph{3GPP
  Technical report, TR 36.814 V9.0.0}, Mar. 2010.

\bibitem{Goldsmith}
A.~Goldsmith and S.-G. Chua, ``Variable-rate variable-power {MQAM} for fading
  channels,'' \emph{{IEEE} Trans. Commun.}, vol.~45, no.~10, pp. 1218--1230,
  Oct 1997.

\end{thebibliography}
\bibliographystyle{IEEEtran}
\end{document}